\documentclass[manuscript]{aastex}

\slugcomment{Submitted to the Astrophysical Journal, 27 September 2013.  Revised 19 December 2013.}

\shorttitle{Carbon Inventory in G328}
\shortauthors{Burton et al.}

\begin{document}


\title{The Carbon Inventory in a \\ Quiescent, Filamentary Molecular Cloud in G328}


\author{Michael G. Burton\altaffilmark{1}, Michael C.B. Ashley, Catherine Braiding, John W.V. Storey}
\affil{School of Physics, University of New South Wales, Sydney, NSW 2052, Australia}

\author{Craig Kulesa}
\affil{Steward Observatory, The University of Arizona, \\ 933 N. Cherry Ave., Tucson, AZ 85721, USA}

\author{David J. Hollenbach}
\affil{Carl Sagan Center, SETI Institute, \\ 189 Bernado Avenue, Mountain View, CA 94043--5203, USA}

\author{Mark Wolfire}
\affil{Astronomy Department, University of Maryland, College Park, MD 20742, USA}

\author{Christian Gl\"uck}
\affil{KOSMA, I. Physikalisches Institut, Universit\"at zu K\"oln, \\ Z\"ulpicher Str.\ 77, 50937 K\"oln, Germany}
\and

\author{Gavin Rowell}
\affil{School of Chemistry and Physics, University of Adelaide, Adelaide, SA 5005, Australia}



\altaffiltext{1}{Email: m.burton@unsw.edu.au}


\begin{abstract}
We present spectral line images of [CI] 809\,GHz, CO J=1--0 115 GHz and HI 1.4\,GHz line emission, and calculate the corresponding C, CO and H column densities, for a sinuous, quiescent Giant Molecular Cloud about 5\,kpc distant along the $l=328^{\circ}$ sightline (hereafter G328) in our Galaxy.  The [CI] data comes from the High Elevation Antarctic Terahertz (HEAT) telescope, a new facility on the summit of the Antarctic plateau where the precipitable water vapor falls to the lowest values found on the surface of the Earth.  The CO and HI datasets come from the Mopra and Parkes/ATCA telescopes, respectively.
We identify a filamentary molecular cloud, $\sim 75 \times 5\,$pc long with mass $\rm \sim 4 \times 10^4\,M_{\odot}$ and a narrow velocity emission range of just 4\,km/s. The morphology and kinematics of this filament are similar in CO, [CI] and HI, though in the latter appears as self-absorption.  We calculate line fluxes and column densities for the three emitting species, which are broadly consistent with a PDR model for a GMC exposed to the average interstellar radiation field.  The  [C/CO] abundance ratio averaged through the filament is found to be approximately unity.  The G328 filament is constrained to be cold ($T_{Dust} < 20$\,K) by the lack of far--IR emission, to show no clear signs of star formation, and to only be mildly turbulent from the narrow line width.  We suggest that it may represent a GMC shortly after formation, or perhaps still be in the process of formation. 
\end{abstract}



\section{Introduction}
\label{sec:intro}
Giant molecular clouds (GMCs) play a fundamental role in the evolution of a spiral galaxy.  They represent the end result of the continual collection of diffuse gas into molecular form in the interstellar medium, some of it having been recycled from past generations of stars.  GMCs are also the sites where the bulk of ongoing star formation in a galaxy then occurs.  They can form from the atomic substrate, as evident in global views of other galaxies \citep[e.g.\ M33 --][]{2003ApJS..149..343E}, or perhaps from the assembly of smaller molecular clouds.  How this occurs within our own Galaxy remains unclear as the process has not yet been seen.  Models range from quasi-static self-gravitational collapse, for instance along magnetic field lines \citep[e.g.][]{2004ASPC..317..248O}, to transient phenomena such as the coalescence of clouds in converging flows in a turbulent medium \citep[e.g.][]{2000A&A...359.1124H, 2012MNRAS.420.1457H}.  Since stars appear to form almost as soon as molecular gas is present, then the rate of formation of GMCs is also instrumental in determining the star formation rate.

Furthermore, despite being molecular, forming clouds will not necessarily be evident through CO line emission, normally the tracer used to delineate their locations.  This is because in the low molecular column densities of clouds forming from the diffuse atomic medium far--UV radiation will dissociate the CO, whereas self-shielding can allow H$_2$ to exist; i.e.\ a pure H$_2$ molecular cloud \citep[sometimes also known as ``dark'' H$_2$ due to the absence of an emitting molecular species;][]{2010ApJ...716.1191W}.  The carbon in such clouds cannot be hidden, however, and will exist primarily in neutral form where it can emit through the [CI] lines at 492 and 809\,GHz (610 \& $371 \mu$m, respectively), or in singly-ionized form, emitting as [CII] 1.90\,THz ($158 \mu$m), or in both.  In either case, its detection is difficult, for observations at these frequencies require exceedingly dry atmospheric conditions or airborne / space platforms.  Facilities able to map these species at good spectral and spatial resolution, and over the large areas of sky that may encompass the environs of forming giant molecular clouds, have not previously been available.  

We present the first data from a new facility at the summit of the Antarctic plateau, Ridge A  \citep{2009PASP..121..976S}. The 62\,cm aperture High Elevation Antarctic Terahertz (HEAT) telescope (Kulesa et al. 2014, in preparation) currently provides spectroscopic imaging in the 810\,GHz lines of CO J=7--6 and [CI] J=2--1, with upcoming capabilities at 1.46\,THz [NII] and 1.90\,THz [CII]\@.  It can already map [CI] emission over significant regions of the Galactic plane, and so be able to search for forming molecular clouds and ``dark'' H$_2$.  The exceedingly low and stable columns of precipitable water vapor at Ridge A (falling below 100$\mu$m for more than 25\% of winter; \citet{2009PASP..121..976S, 2010PASP..122..490Y}) make this possible. We have identified a filament in [CI] in the G328 region that is also evident in molecular (CO) and atomic (HI) survey data taken with similar resolution.  The results of our mapping of this filament are the subject of this paper. [CII] emission has also been identified along a sightline near to this filament by the GOT C+ program using the \textit{Herschel} telescope
\citep{2014A&A}.

The observations are summarised in \S\ref{sec:obs} and the results presented in \S\ref{sec:results}. They are then discussed in \S\ref{sec:discuss} with reference to the physical characteristics of the G328 filament, a PDR model for its line emission, and the carbon abundance within it.

\section{Observations}
\label{sec:obs}
The data presented here comes from three separate spectral line surveys of sub-mm [CI], mm CO and cm HI emission in the southern Galactic plane.  HEAT is imaging the [CI] 809.3\,GHz line ($\rm ^3P_2 -  ^3P_1$) from Ridge A, Antarctica, yielding data cubes with $2'$ and 0.7\,km/s resolution.  $^{12}$CO, $^{13}$CO  and C$^{18}$O J=1--0 emission lines (115.3, 110.2 \& 109.8\,GHz, respectively) were imaged as part of the 22\,m Mopra Southern Galactic Plane CO Survey \citep{2013PASA...30...44B}, with $0.6'$ and 0.1\,km/s resolution.  The 1.420\,GHz HI line comes from archival data in the Parkes--ATCA Southern Galactic Plane Survey \citep[the SGPS;][]{2005ApJS..158..178M} and provides $2'$ and 3\,km/s resolution. The contiguous data set for these species covers $l=328.0$--$328.5^{\circ}, b=-0.4$ to $+0.4^{\circ}$, although the CO and HI images presented here in fact cover a larger region, of $l=328$ to $329^{\circ}, b=-0.5$ to $+0.5^{\circ}$.  [CI] and CO emission extend over a velocity range from $-100$ to 0\,km/s $V_{\rm LSR}$ in the data cube, the HI from $-120$ to $+100$\,km/s. Spatially, the emission from the three species across these velocity ranges covers the entire region mapped.

\section{Results}
\label{sec:results}

A sinuous emission feature was identified in the CO data cubes, extending $\sim 0.8^{\circ} \times 0.05^{\circ}$ in angular size and centred at $-78$\,km/s, with a narrow spectral range of just 4\,km/s.  This filament is analysed in this paper, with an analysis across the full spectral emission range of the cubes presented in Kulesa et al.\ 2014 (in preparation).

\begin{figure}
\hspace*{-0.5cm}\includegraphics[angle=0,scale=0.5]{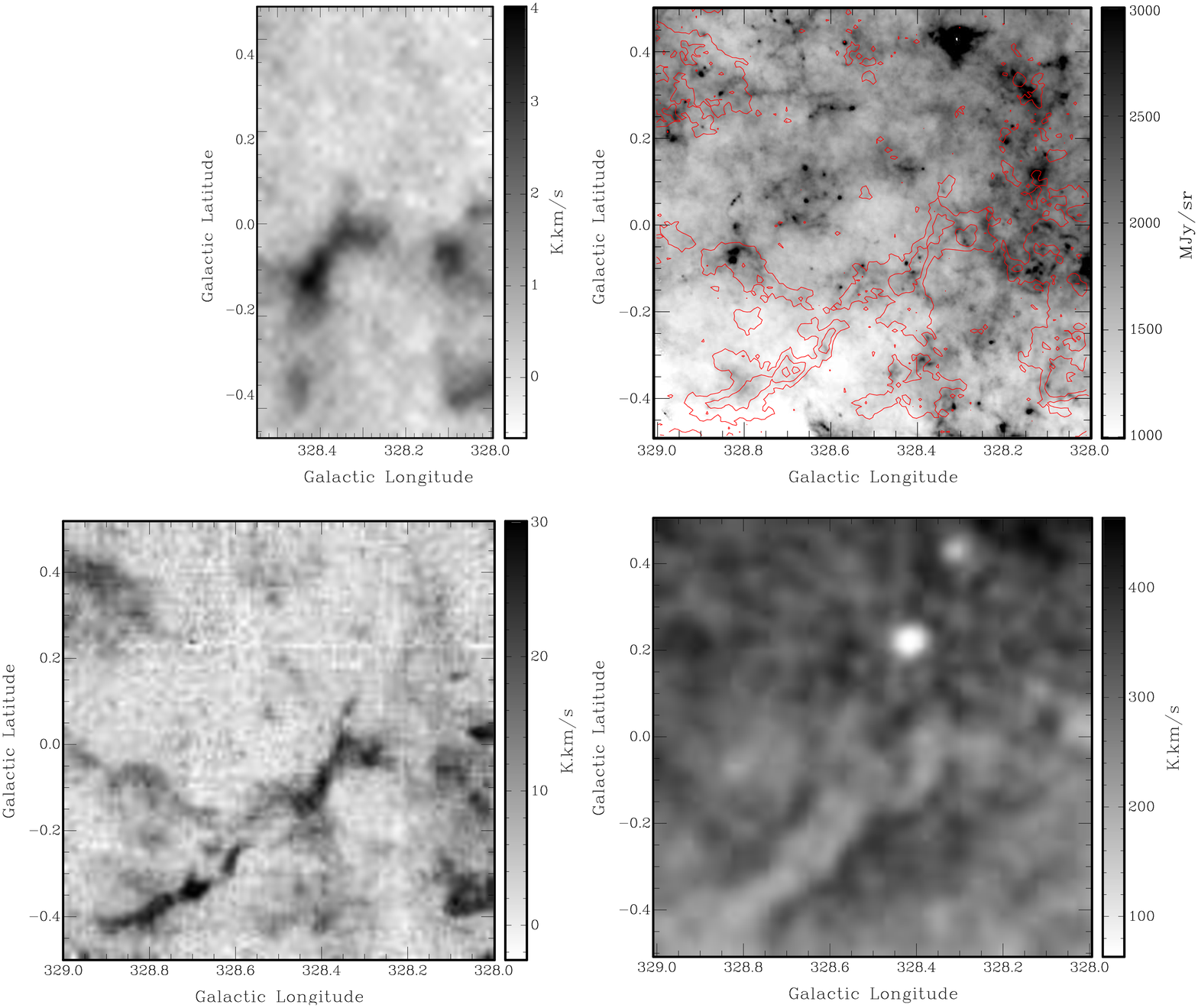}
\caption{Integrated flux images ($T_{MB}$; K km/s) for the G328 region across $-80$ to $-76$ km/s velocity range of the filament.  From top-left, going anti-clockwise: the HEAT [CI] 809 GHz, Mopra $^{12}$CO J=1--0 115\,GHz and Parkes--ATCA 21\,cm HI images. The final image overlays contours of CO on the \textit{Herschel} 250$\mu$m image.  Coordinates are Galactic.  \label{fig:CICOHI}}
\end{figure}

Figure~\ref{fig:CICOHI} presents integrated flux images of the [CI], $^{12}$CO and HI for the G328 filament, together with a 250$\mu$m continuum image of the dust emission from the \textit{Herschel} HiGal program \citep{2010A&A...518L.100M}. The $^{13}$CO image (not shown) looks essentially the same as $^{12}$CO (albeit at $\sim 30\%$ the flux), and is used to determine the CO optical depth.   C$^{18}$O is also detected at the brightest portions of the filament at the $3\sigma$ level, indicative of high optical depths.  It is also notable that the HI is evident as a depression in the level of the extended atomic emission found across the entire G328 region (i.e.\ HI self-absorption or HISA\footnote{Note that the two bright white spots in this image result from absorption of background continuum sources and are apparent at all velocities in the HI data cube.}), whereas the CO and [CI] emission are primarily confined to just the filament itself in this velocity range.  Also notable is that the filament is not apparent in the dust emission map\footnote{A detection of the dust emission from the filament may be limited by the diffuse background level across the \textit{Herschel} Galactic plane images; in the 250, 350 \&  500$\mu$m images this level is $\sim$1,200,  600 \& 200\,MJy/sr, respectively, in G328.}.  There also do not appear to be any IR sources that are clearly associated with the filament, as might be expected if star formation were active in it.

\begin{figure}
\hspace*{-0cm}
\includegraphics[angle=0,scale=0.8]{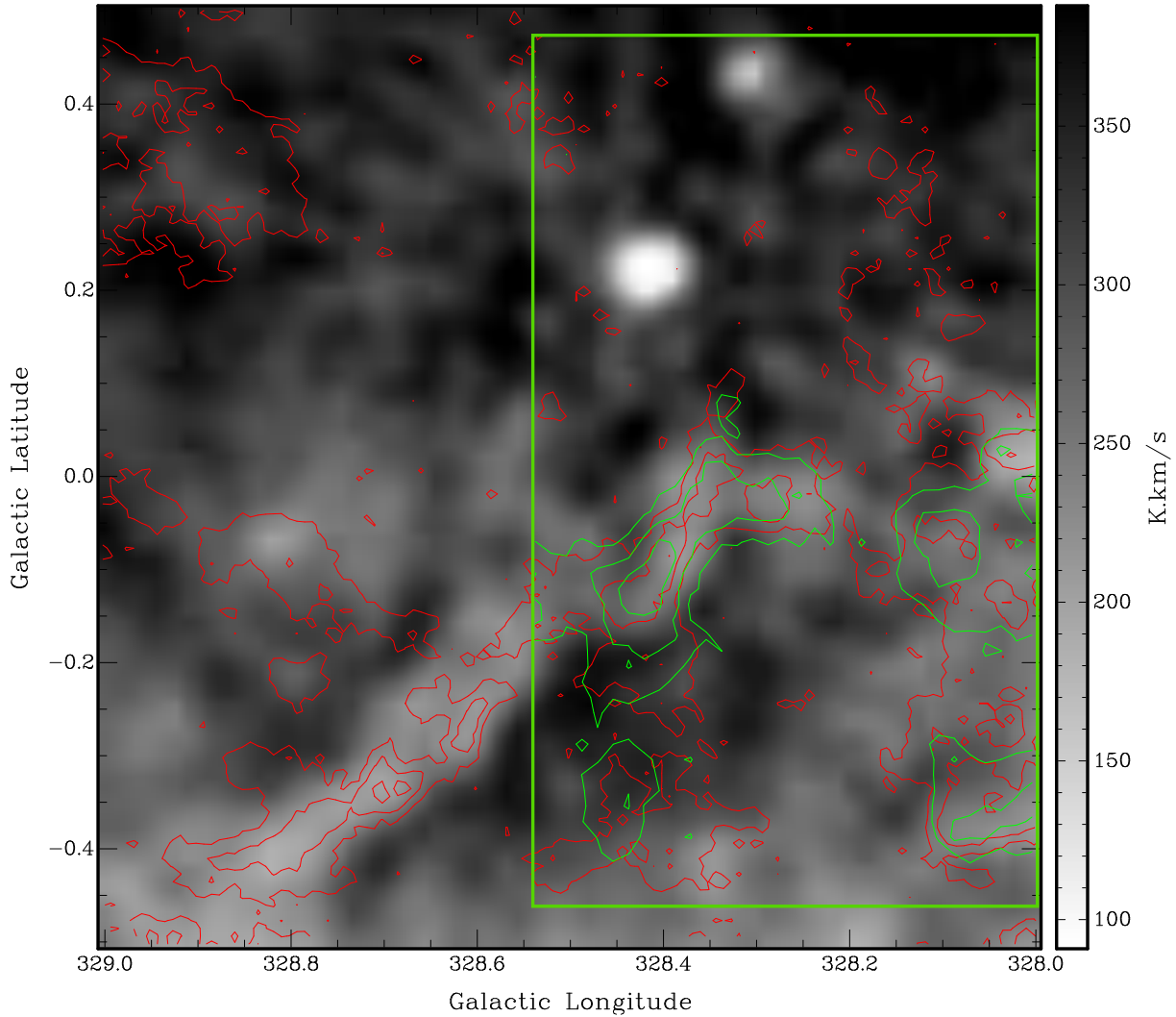}
\caption{Overlay of contours of the [CI] integrated flux (green) and $^{12}$CO (red) on an image of the HI (greyscale; note that white is low and black is high) for the $-80$ to $-76$ km/s velocity range of the G328 filament.  The data for [CI] only extends for $l < 328.5^{\circ}$, as indicated by the green box. The flux scale bar in K km/s is for the HI.  [CI] contours are drawn at 1, 2  \& 3 K km/s; $^{12}$CO contours at 10, 20 \& 30 K km/s (all $T_{MB}$).  \label{fig:contours}}
\end{figure}

Figure~\ref{fig:contours} shows the HI image overlaid with contours of the CO and [CI].    It is clear that these features are closely related.  The HEAT [CI] dataset so far spans only the NW portion of the filament (the ``Head'').  However, given the similarity of the CO and [CI] here, both spectrally and in morphology, it seems likely that the [CI] emission will be co-extensive with the CO along the rest of the filament.

\begin{figure}
\hspace*{-0cm}
\includegraphics[angle=0,scale=0.4]{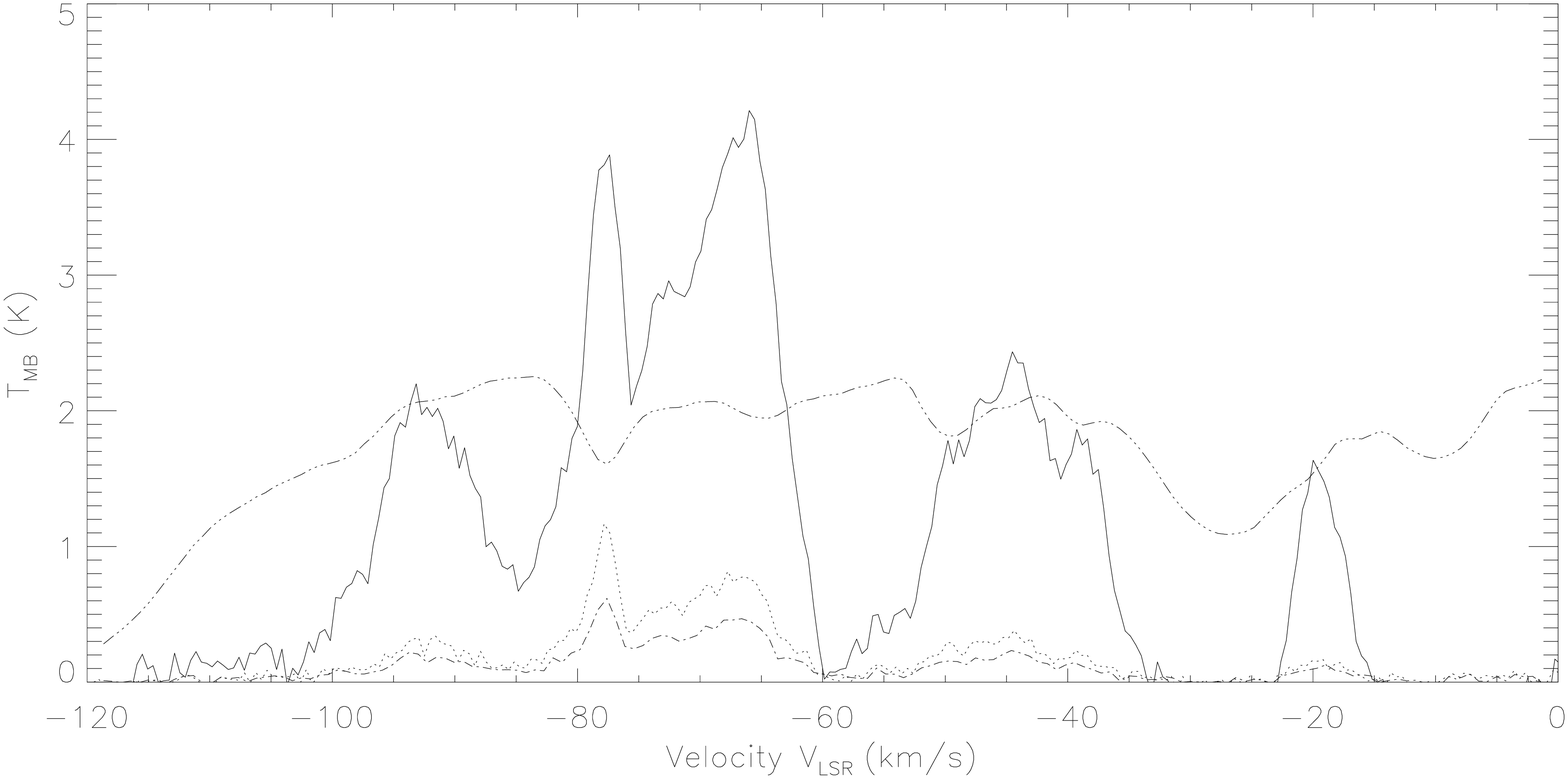} \\
\includegraphics[angle=0,scale=0.4]{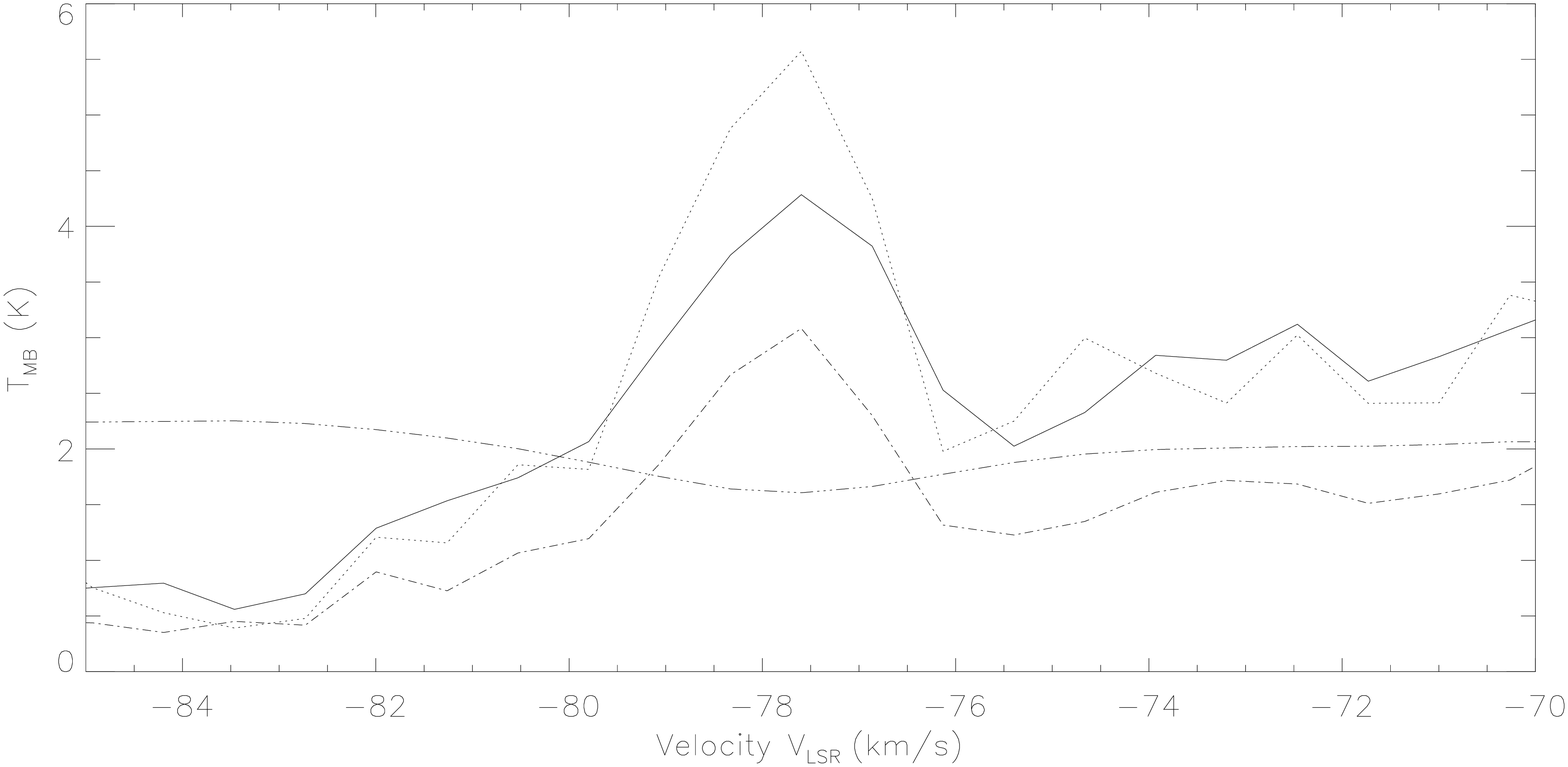}
\caption{Line profiles of the integrated emission from G328 filament for Aperture 1 in Table~\ref{tab:apertures}.  The lines shown are $^{12}$CO (solid), $^{13}$CO (dotted), [CI] (dash-dot) and HI/50 (dash-dot-dot). The top plot shows the entire spectrum at this location, and the bottom plot is zoomed in to show the emission around the -76 to -80\,km\,s$^{-1}$ velocity range analysed in this paper.  In the bottom plot the $^{13}$CO and [CI] lines are also multiplied by a factor 5 for clarity, whereas the HI is divided by 50 in each plot. The HISA is evident as the dip in the HI profile.} \label{fig:profiles}
\end{figure}

\begin{figure}
\hspace*{-0cm}
\includegraphics[scale=0.4]{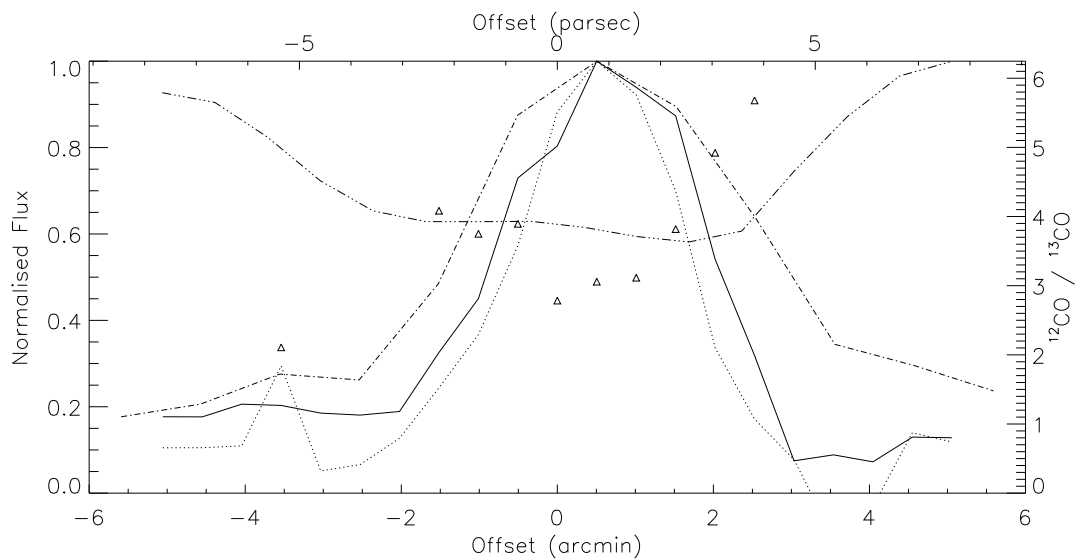}
\caption{Spatial cuts perpendicular to the G328 filament, and normalised to the peak value for each line.  Curves are as follows: $^{12}$CO (solid), $^{13}$CO (dotted), [CI] (dash-dot) and HI (dash-dot-dot). The cut runs from $(l,b) = (328.457,-0.008)$ to $(328.329,-0.118)$, with the offset given in arcminutes on the bottom axis and in parsecs on the top (assuming a source distance of 5.2\,kpc).   The triangles show the $\rm ^{12}CO/^{13}CO$ line ratios across the cut, as indicated on the right-hand axis.  \label{fig:cuts}}
\end{figure}

Figure~\ref{fig:profiles} presents profiles for the lines averaged over the Head of the filament.  We show both the entire spectrum along the sight line as well as zoomed into the velocity range for the G328 filament, for clarity. The CO and [CI] are clearly similar for the filament (i.e.\ same line centre and narrow FWHM), but for HI this is evident as a dip in the profile, which extends considerably beyond the spectral limits shown. Figure~\ref{fig:cuts} shows a spatial cut perpendicular to the filament in the integrated flux images.  The filament is narrow and shows similar structure in CO and [CI].\footnote{We note that if the CO data is smoothed to the spatial resolution of the [CI] then their profiles along this cut are virtually the same.}    As for the velocity profiles, there is a dip in the HI flux across the filament.  The reduction amounts to approx 40\% of the flux level found either side of the filament.

Line parameters are then calculated for 3 different apertures associated with the filament.  The first is for the integrated flux for its Head, and the other two are for the peak pixels seen in the [CI] image.  These apertures are defined in Table~\ref{tab:apertures} and their line fluxes are listed in Table~\ref{tab:fluxes}.  $^{12}$CO optical depths are also calculated from the $^{12}$CO / $^{13}$CO flux ratio, $R$ (i.e. $\tau \sim X/R$ with  $X = [\rm ^{12}C/^{13}C] \sim 50$ appropriate to a Galactocentric radius of $\sim 5$\,kpc; the derived values are only mildly sensitive to its likely variation -- see \citet{2013PASA...30...44B}). The CO optical depths are significant($> 10$) and greatest at the brightest pixels.  

Column densities for CO, C and H are then given in Table~\ref{tab:coldens}.  These are calculated using a standard excitation analysis from the upper level column density given by radiative transfer \citep[e.g.][]{1999ApJ...517..209G, 2013PASA...30...44B}, with standard notation:
\begin{equation}
N_J = T_{MB} \delta V \frac{8 \pi k \nu_{JJ'}^2}{A_{JJ'} h c^3}\frac{\tau}{1 - e^{-\tau}}.
\end{equation}

The column density for the species is then determined from
\begin{equation}
N = \frac{N_J}{g_J} \, Q(T_{ex}) \, e^{T_J/T_{ex}}.
\end{equation}

Here $Q(T)$ is the partition function, taken to be $2 T_{ex}/T_1 \, (T_1 = 5.5$\,K) for CO, 4 for HI, and from fitting to the tabulated values in the Cologne Database for Molecular Spectroscopy (CDMS) for [CI] \citep{2001A&A...370L..49M, 2005JMoSt.742..215M}. The principal assumptions made in these calculations are of a beam filling factor of unity and in the value of the excitation temperature, here taken to be $T_{ex} = 20$\,K\@. $\tau$ is determined for CO as described above and the optically thin limit is assumed for C.\footnote{For the columns that are then determined $\tau_{C}$ can be calculated and is found to be $< 0.2$, so this assumption is self-consistent.} Motivating the temperature assumption, in the PDR model we discuss in \S\ref{sec:pdr} the [CI] and CO lines arise in regions of similar temperature of $\sim 20$\,K\@. Also, combining the [CI] 809\,GHz flux with a low S/N spectrum for the 492\,GHz $^3P_1 - ^3P_0$\,[CI] line obtained with the Nanten2 telescope at the filament's peak (Gl\"{u}ck et al.\ 2014, in preparation) yields a similar value.  Changing $T_{ex}$ by $\pm 10$\,K likewise changes $N({\rm CO})$ by about one-third.   The Boltzmann factor applied to the 60\,K energy level of the 809 GHz [CI] line makes $N({\rm C})$ particularly sensitive to lower values of $T_{ex}$.  For 10\,K it would increase by an order of magnitude; however for 30\,K $N({\rm C})$ falls by only a half. On the other hand, if $T_{ex}$ were significantly less than 20\,K then the Nanten2 492\,GHz flux would have been much higher, given the measured value for the 809\,GHz line. The range in the derived columns for $N({\rm CO})$ and $N({\rm C})$, considering all such likely variations in these parameters, does not alter the essential results that are given below.

The column density of cold atomic gas is estimated from the HI self-absorption profiles following the method of \citet{1978AJ.....83.1607S} and \citet{2000ApJ...540..851G} to determine the absorption optical depth.  We use the difference in fluxes between the bottom of the absorption profile ($T_{\rm min}$) and its value if no absorption had occurred ($T_{\rm interp}$,  obtained by interpolating across the line profile from the values either side) to determine $\tau_{HISA} = {\rm ln}[(pT_{\rm interp}-T_S)/(T_{\rm min} + (p-1)T_{\rm interp} -T_S)]$ (see Table~\ref{tab:fluxes}), where $T_S$ is the spin temperature and $p$ is the fraction of the HI emission assumed to originate from behind the filament (we have also implicitly assumed that the background continuum brightness temperature is very much less than the line brightness temperature in deriving this formula).  We take $T_S = 30$\,K, intermediate between the value in the cold molecular gas ($\sim 20$\,K) and an upper limit equal to the lowest value found in the HI data cube over the entire filament ($\sim 50$\,K).  Following the discussion in \citet{2000ApJ...540..851G}, $p$ is likely to be close to 1 for HISA to be prominent (but note also that it has minimum value given by $p_{\rm low} = 1 + (T_S - T_{\rm min})/T_{\rm interp}$ in the limit that $\tau \rightarrow \infty$; for the values used here this is $\sim 0.6$).   The column of cold atomic gas in the filament is then given by $N{\rm (H_{HISA})} = 1.8 \times 10^{18}  \, T_S \, \tau \, \Delta V\,{\rm cm^{-2}}$ and is listed in Table~\ref{tab:coldens}.  For $T_S =$\,(20K, 50K) these columns should be multiplied by $\sim (0.5, 3)$ times, respectively.  If only two-thirds of the HI emission was from behind the filament (i.e. $p = 2/3$) then the HI columns should be multiplied by a further factor of $\sim 3$.

For C and CO the derived columns are of order a few $\rm \times 10^{17} \,cm^{-2}$.   It is clear that the average C and CO abundances are comparable.  Their ratio is a few times higher than that found in the dense PDR of Carina,  where Nanten2 measurements found $\rm [C/CO] \sim 0.2$ \citep{2008A&A...477..547K}.

For H the total column in emission is found to be $\rm \sim 10^{21} \,cm^{-2}$ in the velocity range of the filament (and so assumed to be at its distance). However this gas is extended over the whole field seen in Fig.~\ref{fig:CICOHI} and so this column cannot be directly associated with atomic gas around the filament, although its projected distance is only 10--20\,pc away. The cold atomic gas is evident as the HISA and has a column estimated to be $\rm \sim 2 \times 10^{20} \, cm^{-2}$, though this figure could readily be increased by $\sim 3$ times, dependent upon the assumed value for $T_S$, the background continuum and particularly the foreground HI contribution, as discussed above.

Along a sightline at $(l,b) = (328.085, 0.0)$ the GOT C+ program with \textit{Herschel} has detected [CII] emission in the same narrow velocity range as the CO, [CI] and HI discussed above \citep{2010A&A...521L..17L, 2014A&A}.  This sightline does not intersect the G328 filament, and lies $\rm \sim 0.3\deg \, (\sim 25\,pc)$ away from its Head, in a region where [CI] emission is absent and the CO emission diffuse and weak (see Fig.~\ref{fig:CICOHI}).  The HI here, however, is seen in emission rather than absorption.  Thus, while this [CII] emission cannot be used to determine the C$^+$ column density associated with the filament, it does suggest that C$^+$ emission is widespread and extends beyond the molecular cloud. The C$^+$ column density at this position is estimated by Langer et al.\ to be several times $\rm 10^{17} \,cm^{-2}$, so comparable in magnitude to that of the CO and C we have derived for the Head. 

\section{Discussion}
\label{sec:discuss}
\subsection{Physical Parameters for the G328 Filament}
\label{sec:parameters}
Striking in the images is the long, sinuous nature of the filament ($\sim 0.8\deg \times 0.05\deg$ in all the spectral features), together with its narrow 4 km/s line width along this entire length.  The central emission velocity ($-76$\,km/s) yields a kinematic distance of (5.2, 9.2)\,kpc (near, far-solutions) and a Galactocentric radius of 5.0\,kpc.  Based on the dip in the HI profile along its length we assume the near-distance.  The line centre is constant to within 1\,km/s along the filament's length.  This yields a size of $\sim 75 \times 5$\,pc, with any velocity gradient or shear along it constrained to be $< 0.02$\,km/s/pc.  This latter limit is comparable to the shear that would produced by Galactic rotation over this distance.

The mean CO line flux over the length of the filament ($\rm \sim 14\,  K \, km/s$ over $0.1 \, {\rm sq.} \deg$) is similar to that over its Head (see Table~\ref{tab:apertures}).   Applying the empirical conversion between flux and column ($\rm X_{CO} = 2.0 \times 10^{20}\, cm^{-2}  / \, [K\, km\, s^{-1}]$; \citet{2013ARA&A..51..207B}) to the integrated flux over the entire filament, and converting to a molecular mass yields $\rm M \sim 4 \times 10^{4}\,M_{\odot}$; i.e.\ comparable to that expected for a low-mass GMC\@.  This implies an average number density $n_{H_2} \sim 7 \times 10^2 \, {\rm cm}^{-3}$ in the filament, given its size (and assuming a depth comparable to its width).  The CO luminosity is $L_{\rm CO\, 1-0} \sim 1 \times 10^4 \, \rm{K \, km \, s^{-1} \, pc^2} \sim 0.1 \,{\rm L_{\sun}}$.


The FWHM of the CO line of 4 km/s is greater than the thermal width for cold gas ($\rm < 1 \, km/s$) so presumably it is mildly turbulent. However any shear along its length is low, being  $\rm < 1$\,km/s over its 75\,pc. The absence of far--IR dust emission in the \textit{Herschel} images associated with the filament implies that the dust must be cold. Assuming grain emissivities, $\kappa_{\nu}$, as given by \citet{1994A&A...291..943O}, then for $T_D \sim 20$\,K the fluxes would be, applying $I_{\nu} = N_{H_2} \mu m_H \kappa_{\nu} R_{DG} B_{\nu}(T_D) \sim$ 1000, 400 \& 150\,MJy/sr at 250, 350 \&  500$\mu$m, respectively. These fluxes are comparable with the Galactic background and so should be discernible in the \textit{Herschel} images.  However, for $T_D \sim 10$\,K they fall to 50--30\,MJy/sr, and so would not readily be distinguished above the background. This implies an upper limit to the dust temperature of $T_D < 20$\,K\@.  The Galactic background in the \textit{Herschel} images can also be used to obtain an upper limit for the dust luminosity; we obtain $L_{\rm Dust} < 2 \times 10^4 \, {\rm L_{\sun}}$, and correspondingly $L/M < 0.5 \, {\rm L_{\sun}/M_{\sun}}$ using the estimated molecular mass. These are strong limits; if the dust emission from the filament was 20\% of the background levels it would likely be discernible, tracing the outline of the filament in the \textit{Herschel} images.  Comparing to the values derived by \citet{1988ApJ...334L..51M} and \cite{1989ApJ...339..149S} for $L/M$ for IR--quiet and non-H~{\small II} region clouds (i.e.\ those lacking in massive star formation), this places the upper limit for the G328 filament at the lower end of the range these authors found, and similarly for the corresponding range found in $L$ vs. $L_{\rm CO}$.  
The G328 filament is thus underluminous compared to IR-quiet GMCs, suggesting that active star formation has not yet commenced. We note that these previous studies used clouds that were bright enough to be seen by \textit{IRAS} against the Galactic background and thus our $L/M$ ratio represents an unbiased and perhaps more representative value for a quiescent GMC\@.

\subsection{PDR Models}
\label{sec:pdr}
We have examined a range of cloud properties using a modified form of the \cite{2010ApJ...716.1191W} photodissociation region code to calculate column densities and line emission to compare with the data.  The model consists of a slab of gas illuminated by the interstellar radiation field on two sides. The gas temperature and the abundances of atomic/molecular species are calculated as a function of optical depth, $A_V$, under the assumptions of thermal balance, chemical equilibrium and constant thermal pressure. We have made a simple approximation to include ${\rm ^{13}CO}$ line transfer by assuming a constant ${\rm ^{12}CO}/{\rm ^{13}CO}$ abundance ratio of 50. For details on the chemistry and thermal processes we refer to \citet{1985ApJ...291..722T}, \citet{2006ApJ...644..283K}, \cite{2010ApJ...716.1191W} and \cite{2012ApJ...754..105H}. While not intended as a comprehensive attempt to fit the data, it has provided a representative model which approximates the observations for Aperture \#2.  The parameters and results  are shown in Table~\ref{tab:pdrmodel}.  

This model adopts an interstellar radiation field of $G_0 = 3$ (with $G_0=1.5$ incident on each side; $G_0$ is the interstellar field strength in units of Habing fields; \citet{1968BAN....19..421H}).   This field is roughly consistent with interstellar fields expected at a Galactocentric radius of 5\,kpc \citep{2003ApJ...587..278W}.   Somewhat surprisingly, the CO and [CI] flux results are not strongly dependent on $G_0$, as discussed in \citet{1993ApJ...402..195W}, because stronger fields drive the C to CO transition deeper into the cloud, where the temperature is lower than at the surface.  Higher fields will increase the C$^+$ 1.9\,THz flux, however. The model shown in Table~\ref{tab:pdrmodel} adopts a primary cosmic ray ionization rate of  $\rm \zeta _{crp}=2\times 10^{-16}\,s^{-1}$, consistent with rates for low column density diffuse clouds determined from observations of H$_3^+$ \citep{2012ApJ...745...91I}, OH$^+$ and H$_3$O$^+$ \citep{2012ApJ...754..105H}.     There is some evidence that the ionization rate is depth dependent in clouds, varying from $\rm \sim 2\times 10^{-16} \,s^{-1}$  in low $A_V$ clouds to values of $\rm \sim 2\times 10^{-17} \,s^{-1}$ in the centers of molecular clouds \citep[e.g.][]{2009A&A...501..619P, 2012ApJ...754..105H, 2012ApJ...745...91I}.   We present in Table~\ref{tab:pdrmodel} a model with the high cosmic ray rate, but an equally good fit to the data can be obtained with the low cosmic ray rate.  We find that the low cosmic ray rate produces a higher fraction of the total CO cooling in the J=1--0 line, because the CO gas is cooler.

The observed CO J=1--0 luminosity of the whole filament is of order 0.1 L$_\odot$.    In the PDR models, this corresponds to a total CO luminosity of the cloud of 3--6\,L$_\odot$, depending on the cosmic ray rate.    For the low cosmic ray rate, the total CO cooling is 3\,L$_\odot$ and most of the heating of the CO is provided by the FUV radiation at intermediate depths into the cloud. For the high cosmic ray  rate the total CO cooling is 6\,L$_\odot$, of which about half the heating is provided by the FUV radiation at intermediate depths and half by cosmic rays distributed through the entire CO region.


The PDR models that fit the data have a total cloud optical depth of $A_V\sim 7$~mags.\ and the interior gas temperature where CO 1--0 becomes optically thick is $\sim 25$ K with an ${\rm H_2}$ density of 650 ${\rm cm^{-3}}$. Note, however, that the model line fluxes are approximately twice that observed indicating that the beam filling factor is $\sim 0.5$. For this beam filling factor, the estimated columns in Table~\ref{tab:coldens} would be multiplied by a factor of $\sim 2$. With this correction,  the model column densities and line fluxes agree with observations to within a factor of 2. We also provide estimates for the C 492 GHz and ${\rm C^+}$ 1.90 THz transitions. The models predict overlapping C$^+$, C and CO emission zones but their peaks are separated by $\Delta A_v \sim 0.5$ mags., equivalent to $\sim 0.5$\,pc, which is not resolvable at the resolution  of the HEAT telescope at 809\,GHz (Fig.~\ref{fig:cuts}), but should begin to be discernible were it at the resolution of the Mopra dataset.  

We also note that the model incident field is $G_0=1.5$.
This field produces dust temperatures at the surface of $\sim 17$ K for silicates and $\sim 23$ K for carbonaceous grains \citep{2011piim.book.....D}. However, the the bulk of radiation field is absorbed, and the dust emission is produced at greater depth, where the field strength is lower. At $\tau_{\rm UV}=1$ the grain temperatures drop to 15 K and 20 K for silicates and carbonaceous grains  respectively. The model dust temperatures are marginally within the limits set by the \textit{Herschel} observations.  

\subsection{Carbon Abundance}
\label{sec:carbon}
Applying the empirical $\rm X_{CO}$ factor to the peak CO flux and comparing to the columns for CO and C yields [C/H] $\sim 8 \times 10^{-5}$, about half  of the measured gas-phase abundance of $1.6 \times 10^{-4}$ \citep{1997ApJ...482L.105S,2004ApJ...605..272S} in diffuse gas. The ${\rm C^+}$ column density is expected to contribute only an additional 10\% to the total. This discrepancy between the estimated carbon abundances from our observations, and those measured in the diffuse gas and used in models, might simply arise from the inadequacy of using the X--factor conversion to provide the total column. 

On the other hand, low carbon abundances (e.g.\ [CO/H] $\sim 4 \times 10^{-5}$) have been previously adopted in analysis of  CO observations \citep[e.g.,][]{2009ApJ...699.1092H, 2010ApJ...723..492R} to match the data, suggesting that most of the non-refractory carbon is hidden is some form as yet undetected.
The model in Table~\ref{tab:pdrmodel} is shown only for representational purposes and is not a best fit from an exhaustive search of model parameter space.  However it demonstrates that reasonable fits can be obtained without depleting the gas phase carbon from the observed diffuse cloud values.  There is no ``missing carbon'' in these models, with the gas phase carbon ($1.6 \times 10^{-4}$ in abundance relative to H nuclei) contained in C$^+$ at the surface, in C at intermediate depth, and then as CO from about $A_V \sim 1$ to the center of the cloud.   Most of the H nuclei are in the CO zone, so the previous observations which ratio the CO column to the H nuclei column and find a few $\times 10^{-5}$ cannot be explained by additional H column in the non-CO zone. Our model explains the observed CO and CI intensities and predicts C$^+$ 1.9\,THz intensities comparable to those observed by GOT C+ in a nearby line of sight \citep{2014A&A}.


The models include freeze-out of atomic carbon and carbon molecules onto grain surfaces using the formalism discussed in \citet{2009ApJ...690.1497H}.  It is possible that some of the CO may be lost from the gas phase due to CO ice formation. For the model presented in Table~\ref{tab:pdrmodel}, at cloud center about 40\% of the carbon is in CO ice. However, the freeze-out zone extends for only a short column in $A_V$ and the column density of CO ice amounts to only $\sim 10$\% of the CO gas-phase column density.


It is likely that there exists and extended, low density atomic halo around the G328 molecular filament, as evident through the HI emission that surrounds it, together with the C$^+$ emission detected along the GOT C+ sightline.  The filament would be expected to connect to the ambient ISM.  However the presence of such an extended halo will have little affect on the PDR model we have presented, as it would not contribute either CO or [CI] line emission.  It would, however, provide low density C$^+$ gas that extends much further from the filament than the constant thermal pressure assumption inherent in our PDR model allows for.  Mapping of the extended [CII] emission will be necessary to address such questions regarding the connection of the molecular to atomic mediums, as well as the possible presence of ``dark'' molecular gas, where CO is not present and the carbon predominantly in the form of C$^+$.



\section{Conclusions}
\label{sec:conclusion}
We have found a quiescent, filamentary molecular cloud in the G328 region, situated about 5\,kpc from the Sun and extending $\sim 75 \times 5$\,pc, with a mass $\rm \sim 4 \times 10^4\,M_{\odot}$, comparable to that of a low mass GMC\@.  The filament is evident in CO, [CI] and in HI self-absorption. The line width is small, $\sim 4$\,km/s along its length, and no velocity gradient is discernible.   Over the sightline through the cloud the average abundance of atomic carbon is similar to that of carbon in its molecular form (i.e.\ CO).  The filament is not discernible in far--IR continuum images, and the flux limits imply that the dust must be cold, $T_D < 20$\,K\@.  The limits on its luminosity place it at the low end of the range in $L/M$ previously found for IR-quiet GMCs. No clear evidence for star formation in the filament can be seen in the IR images, consistent with no significant internal heating.  The molecular cloud is thus cold and has limited internal turbulence.  We speculate that this filament represents a GMC that has recently formed, or indeed may still be in the process of forming.  Star formation has not yet progressed far within it, and there are no, or limited, sources of turbulent energy injection into the gas.  

This hypothesis can be tested through further wide-field THz spectral imaging, in both the [CII] 158$\mu$m (1.90 THz) line and the two [CI] lines. [CII] would be expected to also show a similar structure to the CO and [CI]\@.  As the gas transitions from atomic to molecular form a high ratio of C$^+$ and C to CO is expected, for the external far--UV radiation field will limit the amount of gas that can exist as CO until sufficient column density has been built up to shield it from dissociating radiation.  It would be of particular interest to determine whether the  C and C$^+$ associated with the filament has a more extended distribution than the CO as this scenario would suggest.  It is first planned to upgrade HEAT to operate at 1.9\,THz to measure the extended C$^+$. Then  a 5\,m diameter THz telescope, such as the proposed DATE5 for Dome A in Antarctica,  and the 0.8\,m balloon-borne STO-2, to be launched from McMurdo Station in Antarctica, in late 2015, would provide similar resolution to the CO in the [CI] and [CII] lines, accordingly, allowing one to examine this question further.

\acknowledgments

We thank the referee, Bill Langer, for his positive criticism of this paper.  This has lead to a number of improvements in the text and the tightening of the arguments we have presented.
 
Funding for the HEAT telescope is provided by the National Science Foundation under grant number PLR-0944335.  PLATO-R was funded by Astronomy Australia Limited, as well as the University of New South Wales, as an initiative of the Australian Government being conducted as part of the Super Science Initiative and financed from the Education Investment Fund.  Logistical support for HEAT and PLATO-R is provided by the United States Antarctic Program.  

The Mopra radio telescope is part of the Australia Telescope National Facility.  Operations support was provided by the University of New South Wales and the University of Adelaide.   The UNSW Digital Filter Bank (the MOPS) used for the observations with Mopra  was provided with financial support from the Australian Research Council (ARC), UNSW, Sydney and Monash universities.  We also acknowledge ARC support through Discovery Project DP120101585.  



{\it Facilities:} \facility{Mopra}, \facility{HEAT}, \facility{Parkes}, \facility{ATCA}, \facility{Herschel}.

\clearpage



\clearpage

\clearpage

\clearpage

\clearpage

\clearpage
\begin{deluxetable}{ccrrcccc}
\tablewidth{0pt}
\tablecaption{Apertures \label{tab:apertures}}
\tablehead{
\multicolumn{2}{c}{Aperture} & \colhead{$l$} & \colhead{$b$} & \colhead{$\Delta l$} & \colhead{$\Delta b$} & \colhead{$V_1$} & \colhead{$V_2$} \\
\colhead{Number} & \colhead{Name} & \colhead{$\deg$} & \colhead{$\deg$} & \colhead{\arcmin} & \colhead{\arcmin} & \colhead{km/s} & \colhead{km/s} 
}
\startdata
1 & Filament Head & 328.375 & -0.085 & 15 & 10 & -80 & -76 \\
2 & Peak 1 & 328.420 & -0.120 & 2 & 2 & -80 & -76 \\
3 & Peak 2 & 328.345 & -0.025 & 2 & 2 & -80 & -76 \\
\enddata
\tablecomments{Apertures used for this analysis.  $l$ \& $b$ are the centres for each aperture in Galactic coordinates, $\Delta l$ \& $\Delta b$ their widths and $V_1$ and $V_2$ the velocity limits. 
}
\end{deluxetable}

\clearpage
\begin{deluxetable}{cccccccccc}
\tablewidth{0pt}
\tablecaption{Line Fluxes \label{tab:fluxes}}
\tablehead{
\colhead{Aperture} & \colhead{$^{12}$CO} & \colhead{$^{13}$CO} & \colhead{C$^{18}$O} & \colhead{[CI]} & \colhead{HI} &\colhead{$\tau_{CO}$}  & \colhead{$\rm HI_{min}$} & \colhead{$\rm HI_{interp}$} & \colhead{$\rm \tau_{HISA}$} \\
\colhead{Number} & \colhead{K km/s}   & \colhead{K km/s} & \colhead{K km/s} & \colhead{K km/s} & \colhead{K km/s} && \colhead{K} & \colhead{K}
}
\startdata
1 & $12 \pm 1$ & $3.0 \pm 0.4$ & $< 0.4$           & $1.7 \pm 0.2$ & $380 \pm 10$ & 13 & 58 & 105 & 1.0 \\
2 & $26 \pm 1$ & $9.2 \pm 0.5$ & $1.6 \pm 0.5$ & $4.1 \pm 0.3$ & $360 \pm 10$ & 18 & 67 & 118 & 0.9 \\
3 & $21 \pm 1$ & $6.1 \pm 0.4$ & $0.7 \pm 0.4$ & $3.0 \pm 0.3$ & $340 \pm 10$ & 15 & 63 & 115 & 1.0 \\
\enddata
\tablecomments{Average fluxes calculated over each aperture, $\int T_{MB} dV$, for the 5 lines listed.  The error includes an estimate for the uncertainty of the continuum level as well as the statistical noise in the data.  The $^{12}$CO optical depth is calculated from the $^{12}$CO/$^{13}$CO line ratio, assuming an intrinsic [$^{12}$C/$^{13}$C] ratio of 50. $\rm HI_{min}$ and $\rm HI_{interp}$ are the brightness temperatures in the HI cube at the bottom of the self-absorption profile, and when interpolated across the profile, respectively.  $\rm \tau_{HISA}$ is the derived optical depth due to this self-absorption, assuming $T_S = 30$\,K and $p=1$ (see \S\ref{sec:results}).}
\end{deluxetable}

\clearpage
\begin{deluxetable}{ccccccccccc}
\tablewidth{0pt}
\tablecaption{Column Densities \label{tab:coldens}}
\tablehead{
\colhead{Aperture}  & \colhead{N(CO)}  & \colhead{N(C)} & \colhead{N(H)} & \colhead{$\rm N(H_{HISA})$} & \colhead{[C/CO]}\\
\colhead{Number}  &  \colhead{$\rm 10^{17} cm^{-2}$} &  \colhead{$\rm 10^{17} cm^{-2}$} & \colhead{$\rm 10^{20} cm^{-2}$} & \colhead{$\rm 10^{20} cm^{-2}$} & \colhead{Ratio}
}
\startdata
1  & $1.7 \pm 0.1 $ & $1.2 \pm 0.2 $ & $7.0 \pm 0.1 $ & 2.2  & 0.7 \\
2  & $5.3 \pm 0.2 $ & $2.8 \pm 0.2 $ & $6.6 \pm 0.2 $ & 1.9  & 0.5 \\
3  & $3.5 \pm 0.2 $ & $2.0 \pm 0.2 $ & $6.2 \pm 0.1 $ & 2.1  & 0.6 \\
\enddata
\tablecomments{Column densities in $\rm cm^{-2}$ calculated for each Aperture. $T_{ex} = 20$\,K is assumed for the CO and [CI] analysis, with $\tau$ set equal to the derived value for CO and the optically thin limit for [CI], respectively (see \S\ref{sec:results}).  N(H) is the total column of atomic gas based on the line flux whereas $\rm N(H_{HISA})$ is an estimate for the column of cold atomic hydrogen in the filament from the self-absorption profile with $T_S = 30$\,K and $p=1$, as described in \S\ref{sec:results}.  The former cannot be directly associated with the filament, however and the latter has an uncertainty of a factor a few.  The [C/CO] abundance is the ratio of the relevant column densities.}  
\end{deluxetable}

\clearpage

\clearpage
\begin{deluxetable}{lc}
\tablewidth{0pt}
\tablecaption{PDR Model\label{tab:pdrmodel}}
\tablehead{
\multicolumn{2}{c}{Input Parameters}\\
\colhead{Parameter} & \colhead{Value}}
\startdata
$A_V$ & $\rm 7.2$ mags.  \\
$P_{\rm th} / k$ & $\rm 2.0 \times 10^4$  K  ${\rm cm^{-3}}$ \\
G$_0$ & 3  Habings \\
$\Delta V$ & 2.4 \, km  s$^{-1}$ \\
$\zeta_{\rm crp}$  & $2.0 \times 10^{-16}$ ${\rm s^{-1}}$ \\
$\rm [C/H]$ abundance  & $1.6\times 10^{-4}$ \\
\cutinhead{Model Outputs}
N(H~\small{I}) & $\rm 7.5\times 10^{20} \, cm^{-2}$\\
N(C$^+$) & $\rm 2.4 \times 10^{17} \, cm^{-2}$ \\
N(C) & $\rm 4.7 \times 10^{17} \, cm^{-2}$ \\
N(CO) & $\rm 1.4 \times 10^{18} \, cm^{-2}$ \\
I($^{12}$CO 1--0 115 GHz) & $\rm 51 \, K \, km\, s^{-1}$ \\
I(${\rm ^{13}CO}$ 1--0 110 GHz) & $\rm 15 \, K \, km\, s^{-1}$ \\
I([CI] 492 GHz) & $\rm 21 \, K \, km\, s^{-1}$ \\
I([CI] 810 GHz) & $\rm   9 \, K \, km\, s^{-1}$ \\
I([CII] 1.90 THz) & $\rm 2.4 \, K \, km\, s^{-1}$ \\
$T(\tau_{\rm CO}=1)$ & 25 K\\
$n_{\rm H_2}(\tau_{\rm CO}=1)$ & 650 ${\rm cm^{-3}}$
\enddata
\tablewidth{450pt}
\tablecomments{
Model parameters and predictions for a 2-sided PDR (i.e.\ the model is 1D with radiation incident from both sides).  $A_V=[N({\rm H~I})+2N({\rm H_2})]/2.0\times 10^{21}$ ${\rm cm^{-2}}$ is the optical depth through the cloud. $P_{th}/k$ is the thermal pressure.  $G_0$ is the radiation field strength in free space in units of the Habing field ($\rm 1.6 \times 10^{-3} \, ergs \, s^{-1} \, cm^{-2}$). $\Delta V$  is the Doppler line width in km s$^{-1}$.  $\zeta_{\rm crp}$ is the primary cosmic ray ionization rate per H nucleus. $T(\tau_{\rm CO}=1)$ is the gas temperature and $n_{\rm H_2}(\tau_{\rm CO}=1)$ the ${\rm H_2}$ number density where CO 1--0 becomes optically thick.}
\end{deluxetable}

\end{document}